\documentclass[sigconf]{acmart}
\usepackage{booktabs} 
\usepackage{tikz}
\usetikzlibrary{bayesnet}
\usepackage{algorithm}
\usepackage{multirow}
\usepackage[noend]{algpseudocode}
\usepackage{subfigure}
\usepackage{array}
\setcopyright{rightsretained}

\begin{document}
\copyrightyear{2017} 
\acmYear{2017} 
\setcopyright{acmcopyright}
\acmConference{CIKM'17}{}{November 6--10, 2017, Singapore.}
\acmPrice{15.00}
\acmDOI{https://doi.org/10.1145/3132847.3133023}
\acmISBN{ISBN 978-1-4503-4918-5/17/11}

\title{Unsupervised Extraction of Representative Concepts from Scientific Literature}

\renewcommand*{\thefootnote}{\fnsymbol{footnote}}
\author{Adit Krishnan\footnotemark[1], Aravind Sankar\footnotemark[1], Shi Zhi, Jiawei Han}

\affiliation{%
  \institution{Department of Computer Science}
  \streetaddress{University of Illinois at Urbana-Champaign, USA}
}
\email{{aditk2, asankar3, shizhi2, hanj}@illinois.edu}


%
%
%
%


\begin{abstract}
\stepcounter{footnote}\footnotetext{Equal contribution}
This paper studies the automated categorization and extraction of scientific concepts from titles of scientific articles, in order to gain a deeper understanding of their key contributions and facilitate the construction of a generic academic knowledgebase. Towards this goal, we propose an unsupervised, domain-independent, and scalable two-phase algorithm to type and extract key concept mentions into aspects of interest (e.g., Techniques, Applications, etc.). In the first phase of our algorithm we propose \textit{PhraseType}, a probabilistic generative model which exploits textual features and limited POS tags to broadly segment text snippets into aspect-typed phrases. We extend this model to simultaneously learn aspect-specific features and identify academic domains in multi-domain corpora, since the two tasks mutually enhance each other. In the second phase, we propose an approach based on adaptor grammars to extract fine grained concept mentions from the aspect-typed phrases without the need for any external resources or human effort, in a purely data-driven manner. We apply our technique to study literature from diverse scientific domains and show significant gains over state-of-the-art concept extraction techniques. We also present a qualitative analysis of the results obtained.
\end{abstract}

%
%
%


\pagenumbering{gobble}
\maketitle

\section{Introduction}
In recent times, scientific communities have witnessed dramatic growth in the volume of published literature. This presents the unique opportunity to study the evolution of scientific concepts in the literature, and understand the contributions of scientific articles via their key aspects, such as techniques and applications studied by them. The extracted information could be used to build a general-purpose scientific knowledgebase which can impact a wide range of applications such as discovery of related work, citation recommendation, co-authorship prediction and studying temporal evolution of scientific domains. For instance, construction of a Technique-Application knowledgebase can help answer questions such as - "What methods were developed to solve a particular problem?" and "What were the most popular interdisciplinary techniques or applications in 2016?".

To achieve these objectives, it is necessary to accurately type and extract the key concept mentions that are representative of a scientific article. Titles of publications are often structured to emphasize their most significant contributions. They provide a concise, yet accurate representation of the key concepts studied. Preliminary analysis of a sample from popular computer science venues in the years 1970-2016 indicates that 81\% of all research titles contain atleast two concept mentions, where 73\% of these titles state both techniques and applications and the remaining 27\% contain one of the two aspects. Although a minority may be uninformative, our typing and extraction framework generalizes well to their abstract or introduction texts.

Our problem fundamentally differs from classic \textit{Named Entity Recognition} techniques which focus on natural language text \cite{nlp_entreg} and web resources via distant supervision \cite{distantsupervision}.
Entity phrases corresponding to predefined categories such as person, organization, location etc are detected using trigger words (pvt., corp., ltd., Mr./Mrs. etc.), grammar properties, syntactic structures such as dependency parses, part-of-speech (POS) tagging and textual patterns. In contrast, academic concepts are not associated with consistent trigger words and provide limited syntactic features. Titles lack context and vary in structure and organization. To the best of our knowledge, there is no publicly available up-to-date academic knowledgebase to guide the extraction task. Furthermore, it is hard to generate labeled domain-specific corpora to train supervised NER frameworks on academic text unlike general textual corpora. This makes our problem fundamentally challenging and interesting to solve. The key requirements of our technique are as follows:
\begin{itemize}
\item Independent of supervision via annotated academic text or human curated external resources.
\item Flexible and generalizable to diverse academic domains.
\item Independent of apriori parameters such as length of concept mentions, number of concepts corresponding to each aspect etc.
\end{itemize}
\begin{figure*}[t]
\includegraphics[width=\linewidth]{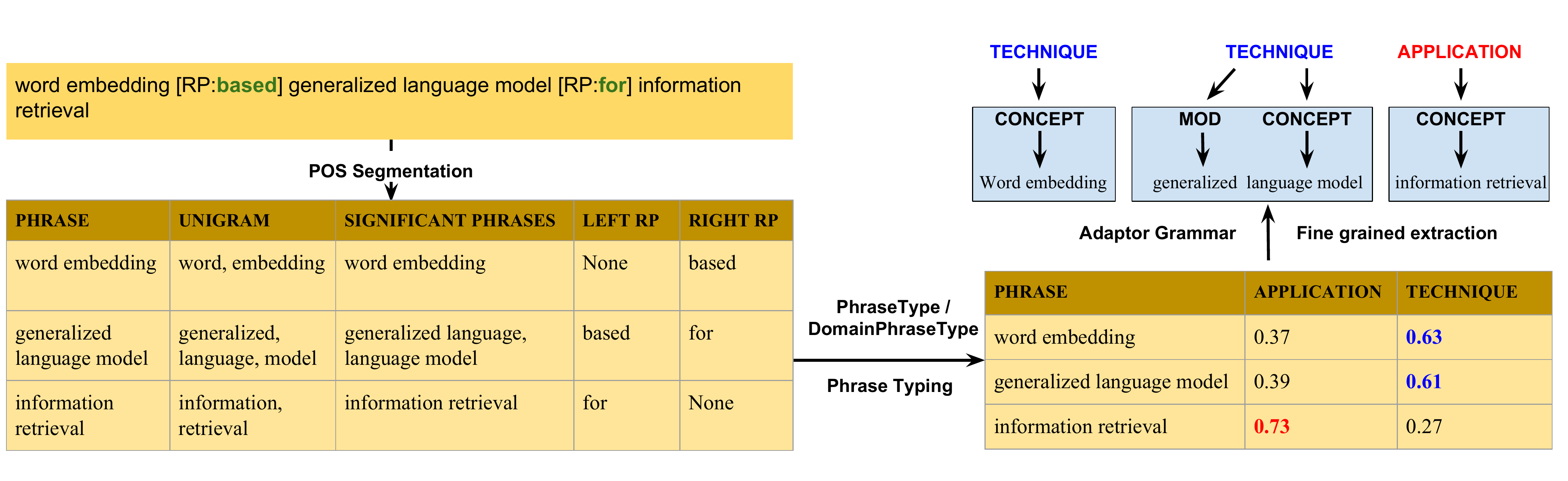}
\vspace{-25pt}
\caption{Pipeline of our concept extraction framework}
\vspace{-5pt}
\end{figure*}
Unlike article text, titles lack contextual information and provide limited textual features rendering conventional NP-chunking \cite{cikm13} and dependency parsing \cite{ijcnlp} based extraction methods ineffective. Previous work in academic concept extraction \cite{cikm13, ijcnlp, facetgist} typically perform aspect or facet typing post extraction of concepts. Alternately, we propose to type phrases rather than individual concept mentions, and subsequently extract concepts from typed phrases. Phrases combine concept mentions such as \textit{tcp} with additional specializing text e.g \textit{improving tcp performance}, which provides greater clarity in aspect-typing the phrase as an application, rather than the \textit{tcp} concept mention. Phrases are structured with connecting relation phrases which can provide insights to their aspect roles, in conjunction with their textual content. Furthermore, aspect typing prior concept extraction provides us the flexibility to impose and learn aspect-specific concept extraction rules.

We thus propose a novel two-step framework that satisfies the above requirements. Our first contribution is an aspect-based generative model \textit{PhraseType} to type phrases by learning representative textual features and the associated relation phrase structure. We also propose a domain-aware extension of our model \textit{DomainPhraseType} by integrating domain identification and aspect inference in a common latent framework. Our second contribution is a data-driven non-parametric rule-based approach to perform fine-grained extraction of concept mentions from aspect-typed phrases, based on \textit{adaptor grammars} \cite{adaptor}. We propose simple grammar rules to parse typed phrases and identify the key concept mentions accurately. The availability of tags from the previous step enables our grammars to learn aspect-specific parses of phrases. 

To the best of our knowledge, ours is the first algorithm that can extract and type concept mentions from academic literature in an unsupervised setting. Our experimental results on over 200,000 multi-domain scientific titles from DBLP and ACL datasets show significant improvements over existing concept extraction techniques in both, typing as well as the quality of extracted concept mentions. We also present qualitative results to establish the utility of extracted concepts and domains.

\section{Problem Definition}
We now define terminology used in this paper and formalize our problem.
\\\
\textbf{Concept:} A concept is a single word or multi-word subphrase (we refer to it as a subphrase to distinguish it from phrases) that represents an academic entity or idea which is of interest to users (i.e it has a meaning and is significant in the corpus), similar to the definitions in \cite{cikm13} and \cite{facetgist}. Concepts are not unique in identity and multiple concepts could refer to the same underlying entity (e.g \textit{DP} and \textit{Dirichlet Process}). \\\
\textbf{Concept Mention:} A concept mention is a specific occurrence or instance of a concept. \\\
\textbf{Aspects:} Users search, read and explore scientific articles via attributes such as techniques, applications etc, which we refer to as aspects. Academic concepts require instance specific aspect typing. \textit{Dirichlet Process} could both, be studied as a problem(Application) as well as proposed as a solution(Technique).\\\
\textbf{Relation Phrase:} A relation phrase denotes a unary or binary relation which associates multiple phrases within a title. Extracting textual relations and applying them to entity typing has been studied in previous work \cite{rp1, rp2}. We use the left and right relation phrases connecting a phrase, as features to perform aspect typing of the phrase.\\\
\textbf{Phrases:} Phrases are contiguous chunks of words separated by relation phrases within a title. Phrases could potentially contain concept mentions and other specializing or modifying words. \\\
\textbf{Modifier:} Modifiers are specializing terms or subphrases that appear in conjunction with concept mentions within phrases. For instance, \textit{Time based} is a \textbf{modifier} for the \textbf{concept mention} \textit{language model} in the \textbf{phrase} \textit{Time based language model} as illustrated in Fig \ref{fig:adaptor_img}.
\label{Problem Definition}
\vspace{-5pt}
\begin{definition} \textbf{Problem Definition:}
\textit{Given an input collection $\mathcal{D}$ of titles of articles, a finite set of aspects $\mathcal{A}$, our goal is to:
\\\textbf{1)} Extract and partition the set of phrases $P$ from $\mathcal{D}$ into $|\mathcal{A}|$ subsets. Each apsect of interest in $\mathcal{A}$ is mapped to one subset of the partition by a mapping $\mathcal{M}$.
\\\textbf{2)} Extract concept mentions and modifiers from each of the $|P|$ aspect-typed phrases. Concept mentions are ascribed the aspect type of the phrase in which they appear.\\}
We achieve the above two goals in two phases of our algorithm, the first phase being \textbf{Phrase Typing} and the second, \textbf{Fine Grained Concept Extraction}. The output of our algorithm is a set of typed concept mentions $\mathcal{C}_{d}$ $\forall$ $d \in \mathcal{D}$ and their corresponding modifier subphrases.
\end{definition}

\section{Phrase Typing}
%
%
%
%
%
In this section, we describe our unsupervised approach to extract and aspect-type scientific phrases.
\subsection{Phrase segmentation}
Input scientific titles are segmented into a set of phrases, and their connecting relation phrases that separate them within the title. We apply part-of-speech tag patterns similar to \cite{facetgist} to identify relation phrases. Additionally, we note here that not every relation phrase is appropriate for segmenting a title. Pointwise Mutual Information(PMI) measure can be applied to the preceding and following words to decide whether to split on a relation phrase or not. This ensures that coherent phrases such as \textit{precision and recall} are not split.

\subsection{PhraseType}
Relation phrases play consistent roles in paper titles and provide strong cues on the aspect role of a candidate phrase. A relation phrase such as \textit{by applying} is likely to link a problem phrase to a solution. However not all titles contain informative relation phrases. Furthermore, we find that 19\% of all titles in our corpus contain no relation phrases. Thus, it is necessary to build a model that combines relation phrases with textual features and learns consistent associations of aspects and text. To this end, we propose a flexible probabilistic generative model \textit{PhraseType} which models the generation of phrases jointly over available evidence.

Each phrase is assumed to be drawn from a single aspect and the corresponding textual features and connecting relation phrases are obtained by sampling from the respective aspect distributions. Aspects are described by their distributions over left and right relation phrases and textual features including unigrams(filtered to remove stop words and words with very low corpus level IDF) and significant multi-word phrases. Significant phrases are defined in a manner similar to \cite{topmine} and extracted at the corpus level. Left and right relation phrases are modeled as separate features to factor associations of the phrase with adjacent phrases.

\begin{figure}[hbtp]
\includegraphics[width=0.5\linewidth]{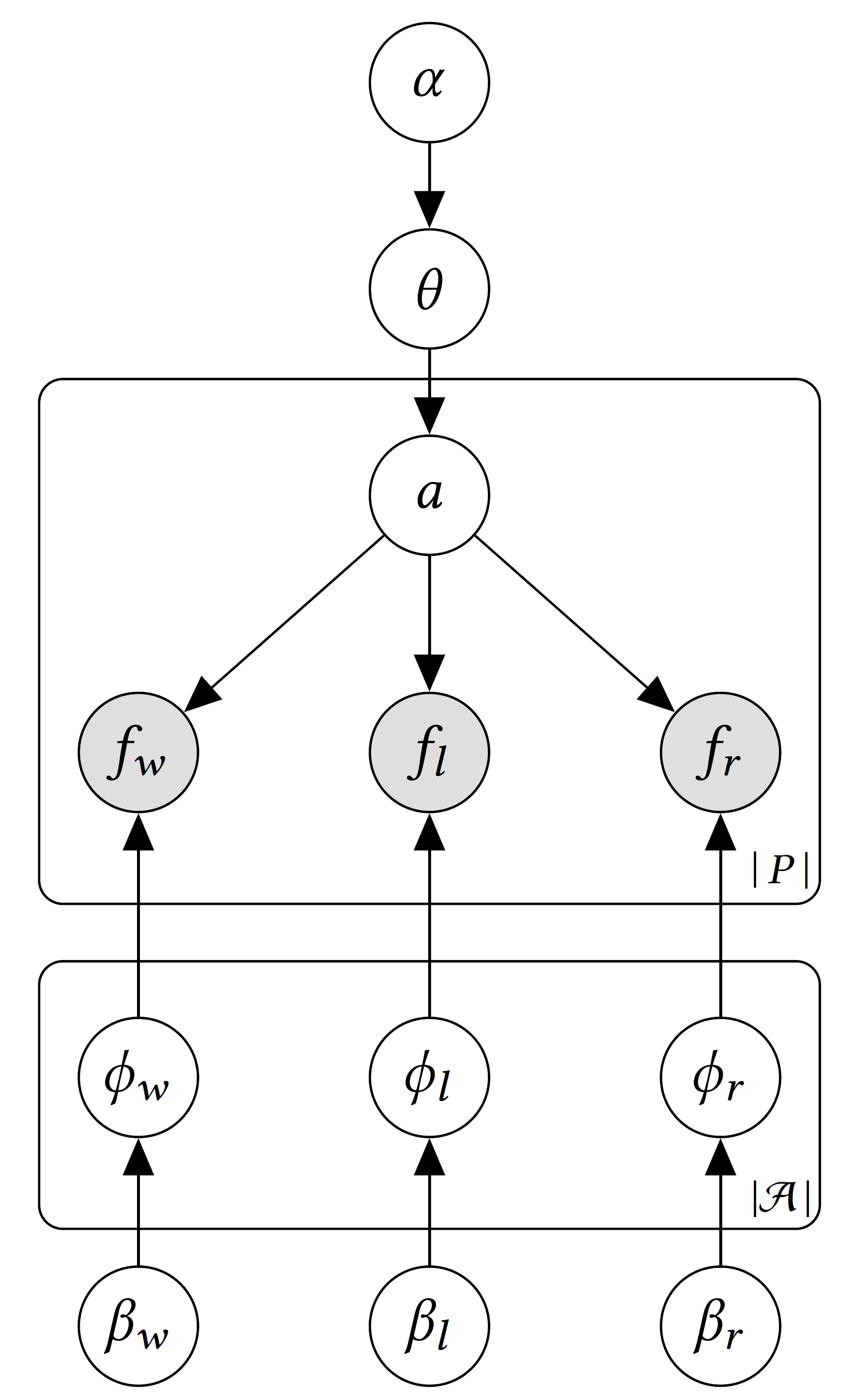}
  \caption{Graphical model for PhraseType}
  \label{phrasetype_model}
  \vspace{-15pt}
\end{figure}

For each phrase $p$ present in the corpus, we choose $p_w$ to denote the set of tokens in $p$, $p_{sp}$ the set of significant phrases in $p$, and $p_{l}$, $p_{r}$ the left and right relation phrases of $p$ respectively. The generative process for a phrase is described in Alg ~\ref{phrasetype} and the corresponding graphical representation in Fig ~\ref{phrasetype_model} (For the sake of brevity we merge $\phi_{sp}$ and $\phi_{w}$ in Fig ~\ref{phrasetype_model}).
\begin{algorithm}[hbtp]
\caption{PhraseType algorithm}
\begin{algorithmic}[1]
\State Draw overall aspect distribution in the corpus, $\theta$ $\sim$ Dir$(\alpha)$
 \For{each aspect $a$}
\State Choose unigram distribution $\phi_{w}^{a}$ $\sim$ $Dir(\beta_{w})$\;
\State Choose significant phrase distribution $\phi_{sp}^{a}$ $\sim$ Dir$(\beta_{w})$\;
\State Choose left relation phrase distribution $\phi_{l}^{a}$ $\sim$ Dir$(\beta_{l})$\;
\State Choose right relation phrase distribution $\phi_{r}^{a}$ $\sim$ Dir$(\beta_{r})$ \;
\EndFor
\For{each phrase $p$}
\State Choose aspect $a \sim$ Mult$(\theta)$
\For{each token $i = 1 ... |p_{w}|$}
\State draw $w_{i}$ $\sim$ Mult$(\phi_{w}^{a})$
\EndFor
\For {each significant phrase $j = 1 ... |p_{sp}|$}
\State draw $sp_{j}$ $\sim$ Mult$(\phi_{sp}^{a})$
\EndFor
\If{$p_l$ exists} draw $p_{l}$ $\sim$ $\phi_{l}^{a}$ \EndIf
 \If{$p_r$ exists} draw $p_{r}$ $\sim$ $\phi_{r}^{a}$ \EndIf
\EndFor
\end{algorithmic}
\label{phrasetype}
\end{algorithm}
\vspace{-10pt}

\subsection{DomainPhraseType}
Most academic domains significantly differ in the scope and content of published work. Modeling aspects at a domain-specific granularity is likely to better disambiguate phrases into appropriate aspects. A simplification could be to use venues directly as domains, however resulting in sparsity issues and not capturing interdisciplinary work well. Most popular venues also contain publications on several themes and diverse tracks. We thus integrate venues and textual features in a common latent framework. This enables us to capture cross-venue similarities and yet provides room to discover diverse intra-venue publications and place them in appropriate domains. To this end, we present \textit{DomainPhraseType} which extends \textit{PhraseType} by factoring domains in the phrase generation process. 

\begin{figure}[hbtp]
\includegraphics[width=0.6\linewidth]{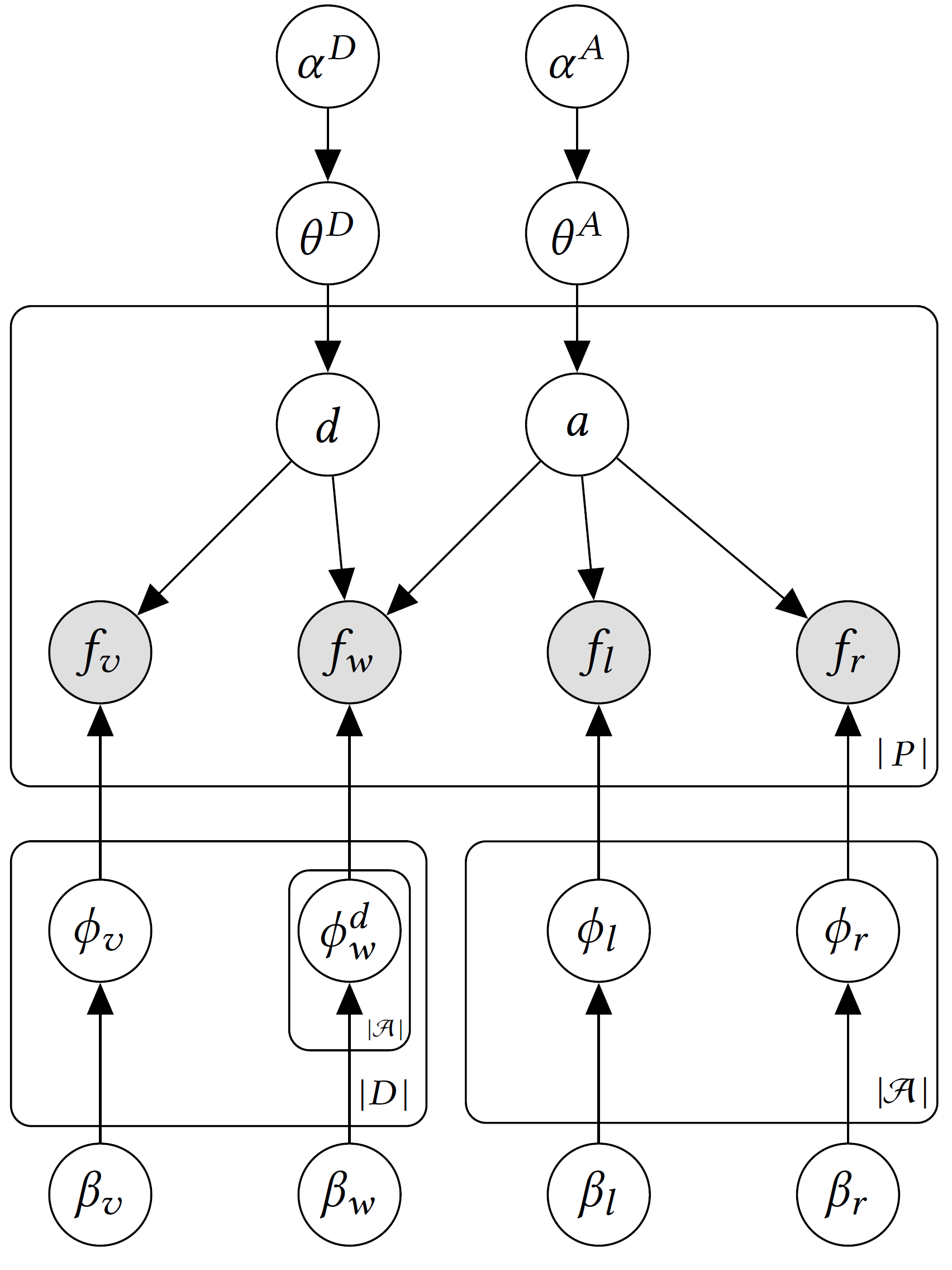}
  \caption{Graphical model for DomainPhraseType}
  \label{domainphrasetype_model}
  \vspace{-10pt}
\end{figure}

To distinguish aspects at a domain-specific granularity, it is necessary to learn textual features specific to a (domain, aspect) pair. Relation phrases however are domain-independent and play a consistent role with respect to different aspects. Additionally, venues often encompass several themes and tracks, although they are fairly indicative of the broad domain of study. Thus, we model domains as simultaneous distributions over aspect-specific textual features, as well as venues. Unlike \textit{PhraseType}, textual features of phrases are now drawn from domain-specific aspect distributions, enabling independent variations in content across domains. The resulting generative process is summarized in Alg ~\ref{domainphrasetype} and the corresponding graphical model in Fig ~\ref{domainphrasetype_model}. Parameter $|D|$ describes the number of domains in the corpus $\mathcal{D}$.
\begin{algorithm}[t]
\caption{DomainPhraseType algorithm}
\begin{algorithmic}[1]

\State Draw overall aspect and domain distributions for the corpus, $\theta^{A}$ $\sim$ Mult$(\alpha^{A})$ and $\theta^{D}$ $\sim$ Mult$(\alpha^{D})$ 
\For{each aspect $a$}
\State Choose left relation phrase distribution, $\phi_{l}^{a}$ $\sim$ $\beta_{l}$
\State Choose right relation phrase distribution, $\phi_{r}^{a}$ $\sim$ $\beta_{r}$
\EndFor
\For{each domain $d$}
\State Draw domain-specific venue distribution $\phi_{v}^{d}$ $\sim$ Dir$(\beta_{v})$
\For{each aspect $a$}
\State Choose unigram distribution $\phi_{w}^{d, a}$ $\sim$ Dir$(\beta_{w})$.
\State Choose significant phrase distribution $\phi_{sp}^{d, a}$ $\sim$ Dir$(\beta_{w})$
\EndFor \EndFor
\For{each phrase $p$}
\State Choose aspect $a \sim$ Mult$(\theta^A)$ and domain $d \sim$ Mult$(\theta^{D})$
\For{each token $i = 1 ... |p_{w}|$}
\State draw $w_{i}$ $\sim$ Mult$(\phi_{w}^{d, a})$
\EndFor
\For{each significant phrase $j = 1 ... |p_{sp}|$}, 
\State draw $sp_{j}$ $\sim$ Mult$(\phi_{sp}^{d, a})$
\EndFor
\State Draw venue $v$ $\sim$ $\phi_{v}^{d}$
\If{$p_l$ exists} draw $p_{l}$ $\sim$ $\phi_{l}^{a}$  
\EndIf
\If{$p_r$ exists} draw $p_{r}$ $\sim$ $\phi_{r}^{a}$ \EndIf
\EndFor
\end{algorithmic}
\label{domainphrasetype}
\end{algorithm}


\subsection{Post-Inference Typing}
In the \textit{PhraseType} model, we compute the posterior distribution over aspects for each phrase as, 
\[ P(a \mid p) \propto P(a) \; P(p_l \mid a) \; P(p_r \mid a) \; \prod_{i = 1}^{|p_{w}|} P(w_i \mid a) \; \prod_{j=1}^{|p_{sp}|} \; P(sp_j \mid a)  \]
and assign it to the most likely aspect. Analogously, in \textit{DomainPhraseType}, we compute the likelihood of (domain, aspect) pairs for each phrase, 
\begin{align*}
 P(d,a \mid p) \; & \propto P(d)  \; P(p_v \mid d) \; P(a) \; P(p_l \mid a) \; P(p_r \mid a)  \\
 & \times \prod_{i = 1}^{|p_{w}|} P(w_i \mid d,a) \; \prod_{j=1}^{|p_{sp}|} \; P(sp_j \mid d,a) 
\end{align*} 
and assign the most likely pair. Phrases with consistently low posteriors across all pairs are discarded.

Additionally we must now map the aspects $a \in [1, |\mathcal{A}|]$ inferred by our model to the aspects of interest, i.e. $\mathcal{A}$ by defining mapping $\mathcal{M}$ from $\mathcal{A}$ to $[1, |\mathcal{A}|]$. Note that there are $|\mathcal{A}|!$ possible ways to do this, however $|\mathcal{A}|$ is a small number in practice. Although our model provides the flexibility to learn any number of aspects, we find that most concept mentions in our datasets are sufficiently differentiated into Techniques and Applications by setting parameter $|\mathcal{A}|$ to 2 in both our models. In other domains such as medical literature, it might be appropriate to learn more than two aspects to partition phrases in medical text.
Let 1 and 2 denote the aspects inferred, and $\mathcal{A}$ = [Technique(T), Application(A)]. We use the distributions $\phi_{l}$ and $\phi_{r}$ of the inferred aspects to set mapping $\mathcal{M}$ either to $M(T,A) = (1,2)$ or $M(T,A) = (2,1)$. Strongly indicative relation phrases such as \textit{by using} and \textit{by applying} are very likely to appear at the left of the Technique phrase of a title, and at the right of the Application phrase. Given a set of indicative relation phrases $\mathcal{RP}$, which are likely to appear as left relation phrases of Technique phrases, and right relation phrases of Application phrases, $\mathcal{M}$ is chosen to maximize the following objective:
$$\mathcal{M} = \arg\max\sum_{rp \in \mathcal{RP}}\hspace{1pt}([\phi_{l}(rp)]_{\mathcal{M(T)}} +  [\phi_{r}(rp)]_{\mathcal{M(A)}})$$ 

\subsection{Temporal Dependencies}
Modeling the temporal evolution of domains is necessary to capture variations that arise over time, in the set of techniques and applications studied by articles published at various venues. To this end, we learn multiple models corresponding to varying time intervals, and explicitly account for expected contiguity in near time-slices. Our objectives with regard to temporal variations are two fold:
\begin{itemize}
\item Sufficient flexibility to describe varying statistical information over different time periods.
\item Smooth evolution of statistical features in a given domain over time.
\end{itemize}

We therefore extend the above models in the time dimension. Our dataset is partitioned into multiple time-slices with roughly the same number of articles. Both models follow the generative processes described above on all phrases in the first time-slice. For subsequent slices the target phrases are modeled in a similar generative manner, however text and venue distributions ($\phi_{sp}^{d, a}, \phi_{w}^{d,a}$ and $\phi_{v}^{d}$) are described by a weighted mixture of the corresponding distributions learned in the previous time-slice, in addition to the prior. This enables us to maintain a connection between domains and aspects learned in different time-slices while also providing flexibility to account for new applications and techniques. Thus $\forall$ T $\geq$ 2:
\begin{itemize}
\item[*] $(\phi_{w}^{d,a})_{t = T}$ $\sim$ $\omega$ $(\phi_{w}^{d,a})_{t = T-1}$ + $(1-\omega)$ Dir($\beta_{w}$)
\item[*] $(\phi_{sp}^{d,a})_{t = T}$ $\sim$ $\omega$ $(\phi_{sp}^{d,a})_{t = T-1}$ + $(1-\omega)$ Dir($\beta_{w}$)
\item[*] $(\phi_{v}^{d})_{t = T}$ $\sim$ $\omega$ $(\phi_{v}^{d})_{t = T-1}$ + $(1-\omega)$ Dir($\beta_{v}$)
\end{itemize}

\section{Fine grained concept extraction}
Academic phrases are most often composed of concepts and modifying subphrases in arbitrary orderings. Concept mentions appear as contiguous units within phrases and are trailed or preceded by modifiers. 
Thus our concept extraction problem can be viewed as shallow parsing or chunking \cite{shallowparsing} of phrases. Unlike grammatical sentences or paragraphs, phrases lack syntactic structure, and the vast majority of them are composed of noun phrases or proper nouns and adjectives. Thus classical chunking models are likely to perform poorly on these phrases.

Unlike generic text fragments, our phrases are most often associated with key atomic concepts which do not display variation in word ordering and always appear as contiguous units across the corpus. For instance, concepts such as \textit{hierarchical clustering} or \textit{peer to peer network} always appear as a single chunk, and are preceded and followed by modifiers e.g. \textit{Incremental hierarchical clustering} or \textit{Analysis of peer to peer network}. This property motivates us to parse phrases with simple rule-based grammars, by statistically discovering concepts in the dataset. 

Probabilistic Context-Free Grammars (PCFGs) are a statistical extension of Context Free Grammars \cite{cfg} that are parametrized with probabilties over production rules, which leads to probability distributions over the possible parses of a phrase. However the independence assumptions render them incapable of learning parses dynamically. Their non-parametric extension, adaptor grammars \cite{adaptor}, can cache parses to learn derivations of phrases in a data-driven manner. Furthermore they are completely unsupervised, which negates the need for any human effort in annotating concepts or training supervised NER frameworks. In the following section, we briefly describe PCFGs and adaptor grammars, and their application to extracting concept mentions and modifiers from phrases. 
\subsection{Probabilistic Context-free Grammars}
A PCFG $\mathcal{G}$ is defined as a quintuple $(N, W ,R, S, \theta)$. Given a finite set of terminals $W$, nonterminals $N$ and start symbol $S$, $\mathcal{G}$ is given by a set of probabilistic grammar rules $(R, \theta)$ where $R$ represents a set of grammar rules while $\theta$ is the set of probabilities associated with each rule.  Let $R_A$ denote the set of all rules that have a nonterminal $A$ in the head position. Each grammar rule $A \rightarrow \beta $ is also called a production and is associated with a corresponding probability 
$\theta_{A \rightarrow \beta}$ which is the probability of expanding the nonterminal $A$ using the production $A \rightarrow \beta$.
According to the definition of a PCFG, we have a normalization constraint for each non-terminal :
$$\sum\limits_{A \rightarrow \beta} \theta_{A \rightarrow \beta \in R_A} = 1 \; \forall A \in N$$ 
The generation of a sentence belonging to the grammar starts from symbol $S$ and each non-terminal is recursively re-written into its derivations according to the probabilistic rules defined by $(R, \theta)$. The rule to be applied at each stage of derivation is chosen independently (of the existing derivation) based on the production probabilities. This results in a hierarchical derivation tree, starting from the start symbol and resulting in a sequence of terminals in the leaf nodes. The final sequence of terminals obtained from the parse tree is called the yield of the derivation tree. A detailed description of PCFGs can be found here \cite{pcfg}.
\subsection{Adaptor Grammars}
\begin{figure}[t]
\includegraphics[width=\linewidth]{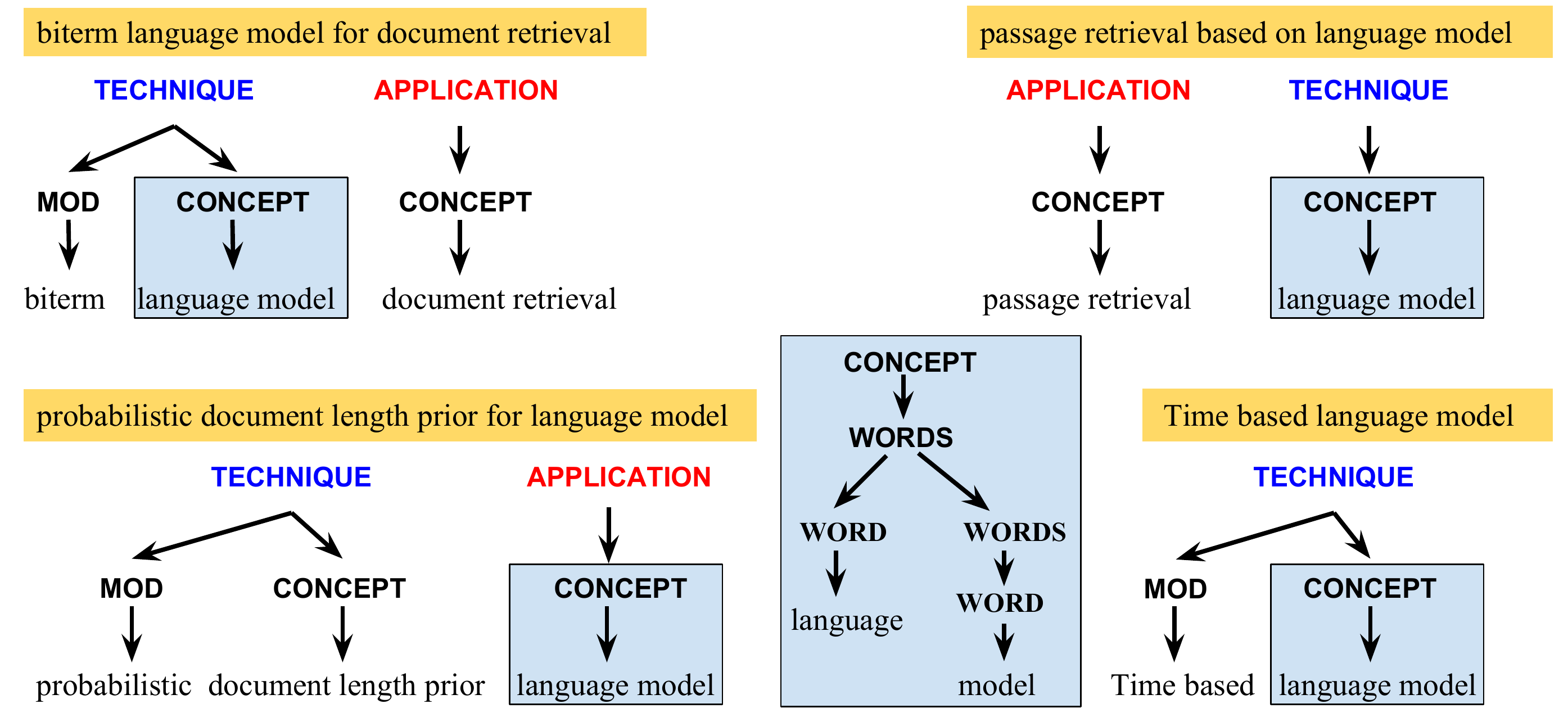}
\caption{Illustration of adapted parse tree involving the adaptor 
\texttt{CONCEPT} to generate the phrase \textit{language model}}
\label{fig:adaptor_img}
\vspace{-11pt}
\end{figure}
PCFGs build derivation trees for each parse independently with a predefined probability on each rule ignoring the yields and structure of previously derived parse trees to decide on rule derivation. For instance, the derivation tree $\textit{Concept} \rightarrow \textit{language model}$ highlighted in Fig ~\ref{fig:adaptor_img}, cannot be learned by a PCFG since every phrase containing \textit{language model} is parsed independently. Adaptor grammars address this by augmenting the probabilistic rules of a PCFG to capture dependencies among successive parses. They jointly model the context and the grammar rules in order to break the independence assumption of PCFGs by caching derivation trees corresponding to previous parses and dynamically expanding the set of derivations in a data-driven fashion.

Concept mentions such as \textit{language model} are likely to appear in several parses and are hence cached by the grammar, which in turn ensures consistent parsing and extraction of the most significant concepts across the corpus. In addition, it has the advantage of being a non-parametric Bayesian model in contrast to PCFG which is parametrized by rule probabilities $\theta$. Adaptor Grammars (Pitman-Yor Grammars) dynamically learn meaningful parse trees for each adapted nonterminal from the data based on the Pitman-Yor process (PYP) \cite{pyp}. Formally, Pitman-Yor Grammar $\mathcal{PYG}$ is defined as,
\vspace{-2pt}
\begin{itemize}
\item Finite set of terminals $W$, nonterminals $N$, rules $R$ and start symbol $S$.
\item Dirichlet prior $\alpha_A$ for the production probabilities $\theta_A$ of each nonterminal $A \in N$, $\theta_A \sim Dir(\alpha_A)$.
\item Set of non-recursive adaptors $C \subseteq N$ with PYP parameters $a_c$, $b_c$ for each adaptor $c \in C$.  
\end{itemize}
 The Chinese Restaurant Process (CRP) \cite{crp} provides a realization of PYP described by a scale parameter, $a$, discount factor $b$ and a base distribution $G_{c}$ for each adaptor $c \in C$. The CRP assumes that dishes are served on an unbounded set of tables, and each customer entering the restaurant decides to either be seated on a pre-occupied table, or a new one. The dishes served on the tables are drawn from the base distribution $G_{c}$. CRP sets up a \textit{rich get richer} dynamics, i.e. new customers are more likely to occupy crowded tables. 
Assume that when the $N^{th}$ customer enters the restaurant, the previous $N-1$ customers labeled $\{1,2,...,N-1\}$ have been seated on $K$ tables ($K \leq N-1$), and the $i^{th}$ customer be seated on table $x_{i} \in \{1,...,K\}$.  The $N^{th}$ customer chooses to sit at $x_{N}$ with the following distribution (note that if he chooses an empty table, this is now the $K+1^{th}$ table),
\vspace{-2pt}
$$P(x_{N} \mid x_{1},...,x_{N-1}) \sim \frac{Kb+a}{N-1+a}\delta_{K+1} + \sum_{k=1}^{K}\frac{m_{k} - b}{N-1+a} \delta_k$$where,
$$m_{k} = \#{x_{i}, i \in \left\lbrace 1,...,N-1 \right\rbrace, x_{i} = k}$$
where $\delta_{K+1}$ refers to the case when a new table is chosen. Thus the customer chooses an occupied table with a probability proportional to the number of occupants ($m_k$) and an unoccupied table proportional to the scale parameter $a$ and the discount factor $b$.  
It can be shown that all customers in CRP are mutually exchangeable and do not alter the distribution. Thus the probability distribution of any sequence of table assignments for customers depends only on the number of customers per table $\mathbf{n} = \{ n_1,..., n_K\}$. This probability is given by,
\begin{equation}
P_{pyp} (\mathbf{n} \mid a, b) = \frac{\prod_{k=1}^K (b(k-1) +a) \prod_{j=1}^{m_k-1} (j-b)}{\prod_{i=0}^{n-1} (i+a)} 
\label{pyp}
\end{equation}
\vspace{-1pt}
where $K$ is the number of occupied tables and ($\sum_{i=1}^{K} n_{i}$) is the total number of customers. In case of a $\mathcal{PYG}$, derivation trees are defined analogous to tables, and customers are instances of adapted non-terminals in the grammar. Thus when a new phrase is parsed, the most likely parse tree assigns the constituent non-terminals in the derivation to the popular tables, hence capturing significant concept mentions in our corpus.

\subsection{Inference}
The objective of inference is to learn a distribution over derivation trees given a collection of phrases as input. Let $P$ be the collection of phrases and $\mathbf{T}$ be the set of derivation trees used to derive $P$. The probability of $\mathbf{T}$ is then given by,
\vspace{-2pt}
\[P (\mathbf{T} \mid \mathbf{\alpha}, \mathbf{a}, \mathbf{b}) = \prod_{A \in N- C} p_{dir} (\mathbf{f_A (T) \mid \alpha_A}) 
\prod_{c \in C} p_{pyp} (\mathbf{n_c(T) \mid a_c, b_c)} \]

where $\mathbf{n_c (T)}$ represents the frequency vector of all adapted rules for adaptor $c$ being observed in $\mathbf{T}$ and $\mathbf{f_A(T)}$ represents the frequency vector of all pcfg rules for nonterminal $A$ being observed in $\mathbf{T}$. Here, $p_{pyp} (\mathbf{n} \mid a,b)$ is as given in Eqn.~\ref{pyp}, while the dirichlet posterior probability $p_{dir} (\mathbf{f} \mid \mathbf{\alpha})$ for a given nonterminal is given by, 
\vspace{-4pt}
\[ p_{dir} (\mathbf{f} \mid \mathbf{\alpha}) = \frac{\Gamma ( \sum\limits_{k=1}^K \alpha_k)}{\Gamma  ( \sum\limits_{k=1}^K  f_k+ \alpha_k)} \prod_{k=1}^K \frac{\Gamma (f_k + \alpha_k)}{\Gamma(\alpha_k)} \]
\vspace{-1pt}
where  $K = |R_A|$ is the number of PCFG rules associated with $A$, and variables $\mathbf{f}$ and $\mathbf{\alpha}$ are both vectors of size $K$.
Given an observed string $x$, in order to compute the posterior distribution over its derivation trees, we need to normalize $p(\mathbf{T} \mid \mathbf{\alpha}, \mathbf{a}, \mathbf{b})$ over all derivation trees that yield $x$. Computing this distribution directly is intractable. We use a MCMC Metropolis-Hastings \cite{adaptor} sampler to perform inference. We refer readers to \cite{adaptor, pcfginference} for a detailed description of MCMC methods for adaptor grammar inference. 
\subsection{Grammar Rules}
\begin{figure}
\includegraphics[width=\linewidth]{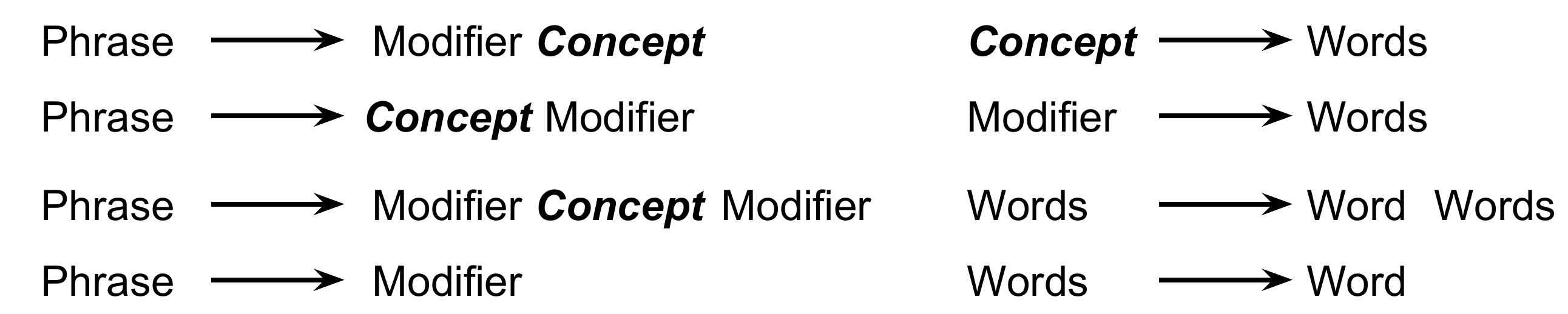}
\caption{Grammar rules to extract concepts and modifiers from typed phrases}
\label{grammar}
\vspace{-12pt}
\end{figure}
The set of phrases $P$ is partitioned by aspect in \textit{PhraseType} and by aspect as well as domain, in case of \textit{DomainPhraseType}. This provides us the flexibility to parse phrases of each aspect (and domain) with a different grammar. Furthermore, parsing each partition separately enables adaptors to recognize discriminative and significant concept mentions specific to each subset which is one of our primary motivations for typing phrases prior to concept extraction. Although a single grammar suffices in the case of Techniques and Application aspects, aspect-specific grammars could also be defined when phrases significantly differ in organization or structure, within our framework.\\
Since phrases are obtained by segmenting titles on relation phrases, it is reasonable to assume that in most cases there is at-most one significant concept mention in a phrase. The set of productions of the adaptor grammar used are illustrated in Fig \ref{grammar} (Adaptor), with Concept being the adapted non-terminal. We also experiment with a variation where both Concept and Mod are adapted (Adaptor:Mod). It appears intuitive to adapt both non-terminals since several modifiers are also statistically significant, such as \textit{high dimensional}, \textit{Analyzing}, \textit{low rank} etc. However, our experimental results appear to indicate that adapting Concept alone performs better. Owing to the structure of the grammar, a competition is set up between Concepts and Mods when both are adapted. This causes a few instances of phrases such as \textit{low rank matrix representation} to be partitioned incorrectly between Mod and Concept causing a mild degradation in performance. When Concept alone is adapted, the most significant subphrase \textit{matrix representation} is extracted as the Concept as expected.\\

\renewcommand*{\thefootnote}{\arabic{footnote}}
\setcounter{footnote}{0}
\section{Experiments}
We evaluate the effectiveness and scalability of our concept extraction framework\footnote{Code: \url{https://github.com/aravindsankar28/Academic-Concept-Extractor}} by conducting experiments on two real-world datasets : DBLP\footnote{DBLP dataset: \url{https://datahub.io/dataset/dblp}} and ACL \cite{acl}.

\subsection{Experimental setup}
79 top conferences were chosen in the DBLP dataset from diverse domains including NLP \& Information Retrieval (IR), Artificial Intelligence and Machine Learning (ML), Databases and Data Mining (DM), Theory and Algorithms (ALG), Compilers and Programming Languages (PL) and Operating Systems \& Computer Networks (NW). The top 50 venues by number of publications were chosen for the ACL dataset. We focus on two primary evaluation tasks.\\\vspace{-10pt}
\begin{flushleft}
\textbf{Quality of concepts:\vspace{-2pt}}
\end{flushleft}
We evaluate the quality of concept mentions identified by each method, without considering the aspect. A set of multi-domain gold standard concepts were chosen from the ACL and DBLP datasets. A random sample of 2,381 documents (for DBLP) and 253 documents (for ACL) containing the chosen gold standard concepts were chosen for evaluation.
\begin{flushleft}
\textbf{Identification of aspect-typed concept mentions:\vspace{-2pt}}
\end{flushleft}
We evaluate the final result set of aspect-typed concept mentions identified by each method on both domain-specific as well as multi-domain corpora. Methods are given credit if both the concept mention as well as the aspect assigned to it are correct.  To perform domain-specific analysis, we manually partition the set of titles in the DBLP dataset into 6 categories based on the venues, and use the unpartitioned DBLP and ACL datasets directly for multi-domain experiments.\\\

\begin{table}[hbtp]
\begin{tabular}{lll}
\toprule
\textbf{Dataset} & \textbf{DBLP} & \textbf{ACL} \\ 
\midrule
Titles & 188974 & 14840 \\ 
Venues & 79 & 50 \\ 
Gold Standard titles  & 740 & 100 \\ 
Gold Standard Technique & 630 &  96\\ 
Gold Standard Application & 783 & 108 \\ 
\bottomrule
\end{tabular}
\caption{Dataset and Gold Standard statistics}
\label{dataset1}
\vspace{-20pt}
\end{table}
A subset of titles in each dataset were annotated with typed concept mentions appearing in their text. Each concept mention was identified and typed to the most appropriate aspect among Technique and Application independently by a pair of experts. The inter-annotator agreement (kappa-value) was found to be 0.86 on DBLP and 0.93 on ACL and the titles where the annotators agreed were chosen for evaluation. Table \ref{dataset1} summarizes the details of corpus and gold standard annotations. Our gold-standard annotations are publicly available online\footnote{\url{https://sites.google.com/site/conceptextraction2/}}.
\textbf{\\\\Evaluation Metrics:\\}
For \textbf{concept quality evaluation}, we compute the F1 score with Precision and Recall. Precision is computed as the ratio of correctly identified concept mentions to the total number of identified mentions. Recall is defined as the ratio of correctly identified concept mentions to the total number of mentions of gold standard concepts in the chosen subset of documents.\\
For \textbf{identification of typed concept mentions}, precision is defined as the ratio of correctly identified and typed concept mentions to the total number of identified mentions. Recall is defined as the ratio of correctly identified and typed concept mentions to the total number of typed concept mentions chosen by the experts.  \\
\renewcommand*{\thefootnote}{\fnsymbol{footnote}}

\begin{flushleft}
\vspace{-10pt}
\textbf{Baselines:\\}
\end{flushleft}
\vspace{-2pt}To evaluate concept quality, we compare against two mention extraction techniques  in literature - Shallow parsing and Phrase Segmentation. Specifically, we compare against : 1) Noun Phrase (NP) chunking and 2) SegPhrase \cite{segphrase}.
To evaluate identification of aspect-typed concept mentions, we compare our algorithms with multiple strong baselines: 
\begin{itemize}
\item \textbf{Bootstrapping + NP chunking} \cite{cikm13} : This is a bootstrapping based concept extraction approach and is currently the state-of-the-art technique for concept extraction in scientific literature.
\item  \textbf{Bootstrapping + Segphrase} : We use a phrase-segmentation algorithm Segphrase \cite{segphrase} to generate candidate concept mentions and apply the above bootstrapping algorithm to extract typed concepts. 
\item \textbf{PhraseType + PCFG}: We use PhraseType combined with a PCFG grammar to extract aspect-typed concepts.
\item \textbf{PhraseType + Adaptor}:  This uses our PhraseType model to extract aspect-typed phrases and performs concept extraction using the Adaptor grammar defined in Fig \ref{grammar} with Concept being adapted.
\item \textbf{DomainPhraseType +Adaptor}: This uses DomainPhraseType to extract aspect-typed phrases and performs concept extraction independently for each domain using the productions defined in Fig \ref{grammar} with Concept being adapted.
\item \textbf{DomainPhraseType +Adaptor:Mod}: This uses DomainPhraseType as above and performs concept extraction using the productions defined in Fig \ref{grammar} while adapting both Mod and Concept non-terminals.
\end{itemize}
For the bootstrapping algorithms, we use a development set of 20 titles in each dataset and set the parameters $(k,n,t)$ to $(2000, 200, 2)$ as recommended in \cite{cikm13}. For PhraseType, we set parameters $\alpha = 50/|\mathcal{A}|$ and $\beta_w = \beta_l = \beta_r = 0.01$, while for DomainPhraseType, we set $\alpha_A = 50/ |\mathcal{A}|$ , $\alpha_D = 50/|D|$ and $\beta_w = \beta_l = \beta_r = \beta_v =  0.01$ and perform inferencing with collapsed gibbs sampling. Temporal parameter $\omega$ was set to 0.5. In our experiments, we run mcmc samplers for 1000 iterations. For \textit{DomainPhraseType}, we varied the number of domains for each dataset and found that $|D| = 10$ in DBLP and $|D| = 5$ in ACL result in the best F1-scores (Fig. \ref{performance_domains}). Discount and scale parameters of adaptors $(a,b)$ were set to $(0.5, 0.5)$ in both \textit{Adaptor} and \textit{Adaptor:Mod} and dirichlet prior $\alpha_{A}$ is set to 0.01. 
\begin{table}[H] \footnotesize
\vspace{-7.5pt}
\begin{tabular}{@{}p{0.25\linewidth}p{0.09\linewidth}p{0.09\linewidth}p{0.09\linewidth}p{0.09\linewidth}p{0.09\linewidth}p{0.09\linewidth}}
\toprule
\multirow{2}{*}{\textbf{Method \textbackslash Dataset}} & \multicolumn{3}{c}{\textbf{DBLP}}  &   \multicolumn{3}{c}{\textbf{ACL} } \\ 
\cmidrule(lr){2-4} \cmidrule(lr){5-7}
 & \textbf{Prec} & \textbf{Rec} & \textbf{F1} & \textbf{Prec} & \textbf{Rec} & \textbf{F1}\\ 
\midrule 
NP chunking & 0.483 & 0.292 & 0.364 & 0.509 & 0.279 & 0.360\\ 
SegPhrase & 0.652 & 0.376 & 0.477 & 0.784 & 0.451 & 0.573\\
PhraseType + Adaptor & \textbf{0.699} & \textbf{0.739} & \textbf{0.718} & \textbf{0.806} & \textbf{0.731} & \textbf{0.767}\\ 
DomainPhraseType + Adaptor:Mod & 0.623 & 0.644 & 0.633 & 0.732 & 0.694 & 0.713 \\
DomainPhraseType + Adaptor & 0.698 & 0.736 & 0.716 & 0.757 & 0.709 & 0.732 \\ 
\bottomrule
\end{tabular}
\caption{Concept quality performance comparison with baselines on DBLP and ACL}
\label{concept_quality}
\end{table}
\vspace{-20pt}

\subsection{Experimental Results}

\begin{table*}[t]\small
\begin{tabular}{@{}p{0.25\linewidth}p{0.045\linewidth}p{0.045\linewidth}p{0.08\linewidth}p{0.045\linewidth}p{0.045\linewidth}p{0.08\linewidth}p{0.045\linewidth}p{0.045\linewidth}p{0.06\linewidth}@{}}
\toprule
\textbf{Method\textbackslash Domain} & \multicolumn{3}{c}{\textbf{IR}}  &   \multicolumn{3}{c}{\textbf{ML} }  &  \multicolumn{3}{c}{\textbf{DM}} \\ 
\cmidrule(r{4pt}){2-4} \cmidrule(l){5-7} \cmidrule(l){8-10}
& Prec & Rec  & F1-Score & Prec & Rec  & F1-Score& Prec & Rec  & F1-Score\\
\midrule
Bootstrapping + NP & 0.437 & 0.325 & 0.373 & 0.4375 & 0.307 & 0.361 & 0.382 & 0.240 & 0.295\\ 
Bootstrapping + Segphrase & \textbf{0.717} & 0.497 & 0.587 & 0.280 & 0.203 & 0.235 & 0.583 & 0.440 & 0.502\\ 
PhraseType + PCFG & 0.444 & 0.487 & 0.465 & 0.374 & 0.390 & 0.382 & 0.364 & 0.434 & 0.396\\ 
PhraseType + Adaptor:Mod & 0.599 & 0.669 & 0.632 & 0.513 & 0.522 & 0.517 & 0.537 & 0.657 & 0.591\\  
PhraseType + Adaptor & 0.712 & \textbf{0.793} & \textbf{0.750} & \textbf{0.653} & \textbf{0.681} & \textbf{0.667} & \textbf{0.584} & \textbf{0.714} & \textbf{0.642}\\  
\midrule
 & \multicolumn{3}{c}{\textbf{PL}}  &   \multicolumn{3}{c}{\textbf{ALG} }  &  \multicolumn{3}{c}{\textbf{NW}} \\ 
\midrule
Bootstrapping + NP & 0.548 & 0.398 & 0.461 & 0.376 & 0.244 & 0.296 & 0.344 & 0.297 & 0.319\\ 
Bootstrapping + Segphrase & \textbf{0.617} & 0.425 & 0.503 & 0.518 & 0.359 & 0.424 & 0.253	& 0.227	& 0.239\\  
PhraseType + PCFG & 0.478 & 0.478 & 0.478 & 0.378 & 0.436 & 0.405 & 0.145 & 0.158 & 0.151 \\ 
PhraseType + Adaptor:Mod & 0.576 & 0.569 & 0.572 & 0.506 & 0.583 & 0.542 & 0.402 & 0.445 & 0.422\\ 
PhraseType + Adaptor & 0.604 & \textbf{0.607} & \textbf{0.605} & \textbf{0.560} & \textbf{0.654} & \textbf{0.603} & \textbf{0.557} & \textbf{0.623} & \textbf{0.588}\\ 
\bottomrule
\end{tabular}
\vspace{2pt}
\caption{DBLP : Domain-specific results (Precision, Recall and F1 scores)  - comparing PhraseType with baselines}
\label{dblp_domains}
\vspace{-15pt}
\end{table*}

\begin{center}
\begin{table*}[t]\small
\begin{tabular}{@{}p{0.05\linewidth}p{0.25\linewidth}p{0.045\linewidth}p{0.045\linewidth}p{0.08\linewidth}p{0.045\linewidth}p{0.045\linewidth}p{0.08\linewidth}p{0.045\linewidth}p{0.045\linewidth}p{0.06\linewidth}@{}}
\toprule
\multirow{2}{*}{\textbf{Dataset}} & \multirow{2}{*}{\textbf{Method}} & \multicolumn{3}{c}{\textbf{Application}} & \multicolumn{3}{c}{\textbf{Technique}} & \multicolumn{3}{c}{\textbf{Overall}} \\
\cmidrule(r{4pt}){3-5} \cmidrule(l){6-8} \cmidrule(l){9-11}
& & Prec & Rec & F1-Score& Prec & Rec & F1-Score& Prec & Rec & F1-Score \\
\midrule
\multirow{7}{*}{\textbf{DBLP}} & Bootstrapping + NP & 0.330 & 0.323 & 0.326 & 0.424 & 0.082 & 0.137 & 0.338 & 0.213 & 0.261\\ 
& Bootstrapping + Segphrase & 0.418 & 0.432 & 0.425 & 0.431 & 0.053 & 0.094 & 0.419 & 0.253 & 0.316\\ 
& PhraseType + PCFG & 0.369 & 0.381 & 0.375 & 0.370 & 0.425 & 0.396 & 0.370 & 0.402 & 0.385\\ 
& PhraseType + Adaptor & 0.604 & 0.628 & 0.616 & 0.554 & 0.653 & 0.599 & 0.578 & 0.640 & 0.607\\
& DomainPhraseType + PCFG & 0.412 & 0.430 & 0.421 & 0.397 & 0.456 & 0.424 & 0.405 & 0.443 & 0.423\\
& DomainPhraseType + Adaptor:Mod & 0.603 & 0.618 & 0.610 & 0.523 & 0.598 & 0.558 & 0.563 & 0.609 & 0.585\\
& DomainPhraseType + Adaptor & \textbf{0.657} & \textbf{0.692} & \textbf{0.674} & \textbf{0.595} & \textbf{0.689} & \textbf{0.639} & \textbf{0.623} & \textbf{0.691} & \textbf{0.655} \\
\midrule
\multirow{7}{*}{\textbf{ACL}} & Bootstrapping + NP & 0.283 & 0.265 & 0.274 & 0.500 & 0.079 & 0.136 & 0.311 & 0.177 & 0.226\\
& Bootstrapping + Segphrase & 0.655 & 0.582 & 0.616 & 0.625 & 0.170 & 0.267 & 0.648 & 0.387 & 0.485\\ 
& PhraseType + PCFG & 0.326 & 0.316 & 0.321 & 0.341 & 0.341 & 0.341 & 0.333 & 0.328 & 0.330\\  
& PhraseType + Adaptor & 0.645 & 0.612 & 0.628 & 0.561 & 0.522 & 0.541 & 0.606 & 0.569 & 0.587\\ 
& DomainPhraseType + PCFG & 0.412 & 0.408 & 0.410 & 0.413 & 0.375 & 0.393 & 0.412 & 0.392 & 0.402\\  
& DomainPhraseType + Adaptor:Mod & 0.680 & 0.673 & 0.676 & 0.616 & \textbf{0.602} & \textbf{0.609} & 0.650 & 0.639 & 0.645\\ 
& DomainPhraseType + Adaptor & \textbf{0.730} & \textbf{0.745} & \textbf{0.737} & \textbf{0.629} & 0.579 & 0.603 & \textbf{0.685} & \textbf{0.667} & \textbf{0.676} \\ 
\bottomrule
\end{tabular}
\caption{DBLP, ACL: Precision, Recall and F1 scores  - Performance comparisons with baselines on individual aspects}
\label{dblp_aspect}
\vspace{-15pt}
\end{table*}

\end{center}
\vspace{-10pt}
\textbf{Quality of concepts:} As depicted in Table \ref{concept_quality}, the concept extraction techniques based on adaptor grammars indicate a significant performance gain over other baselines on both datasets. Adaptor grammars exploit corpus-level statistics to accurately identify the key concept mentions in each phrase which leads to better quality concept mentions in comparison to shallow parsing and phrase segmentation. Amongst the baselines, we find \textit{SegPhrase} to have a high precision since it extracts only high quality phrases from the titles while all of them suffer from poor recall due to their inability to extract fine-grained concept mentions accurately. 

\begin{figure*}
\subfigure[b][]{
        \centering
        \includegraphics[width=0.32\linewidth]{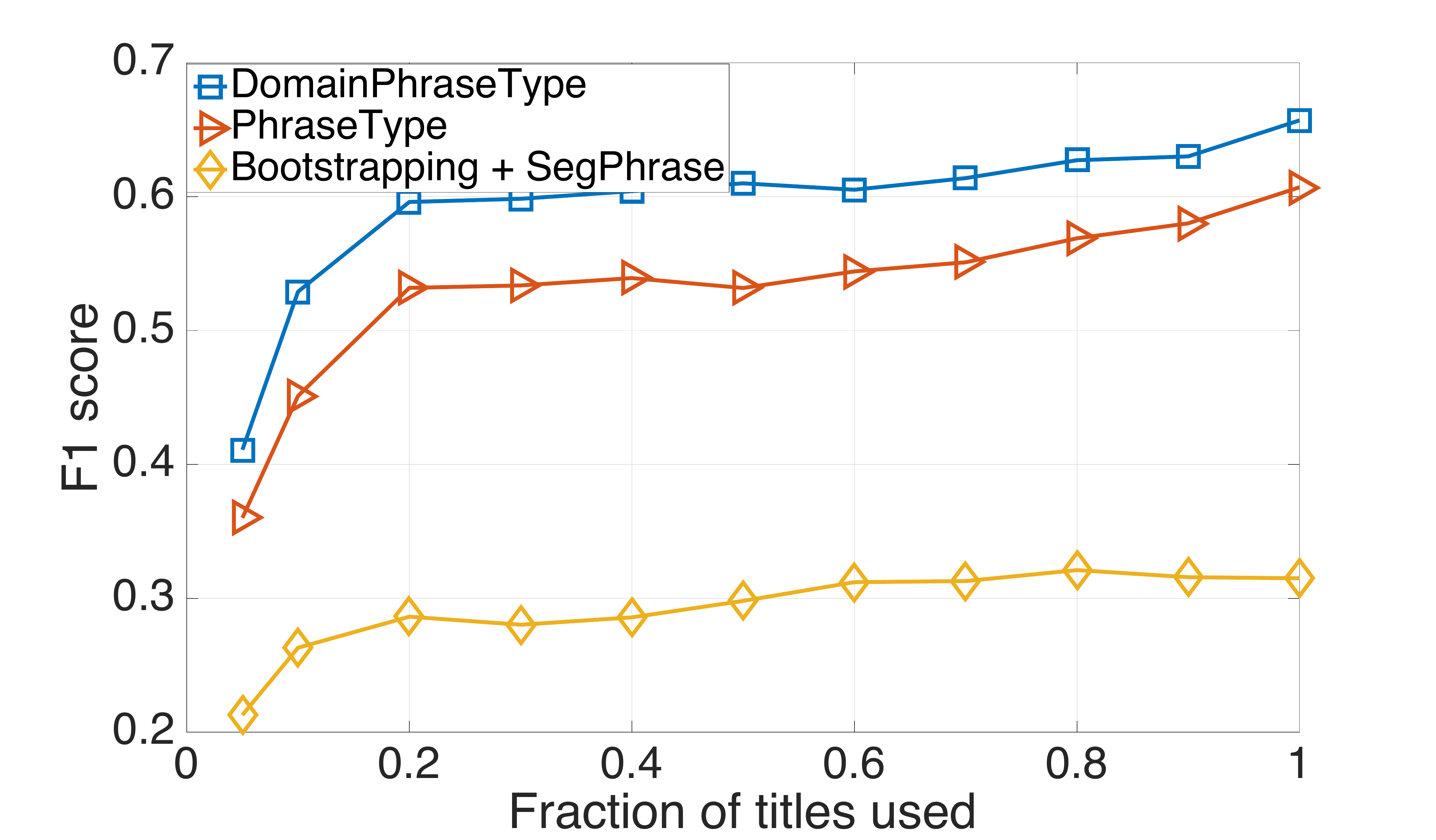}
        \label{performance_corpus}}
\subfigure[b][]{
        \centering
        \includegraphics[width=0.32\linewidth]{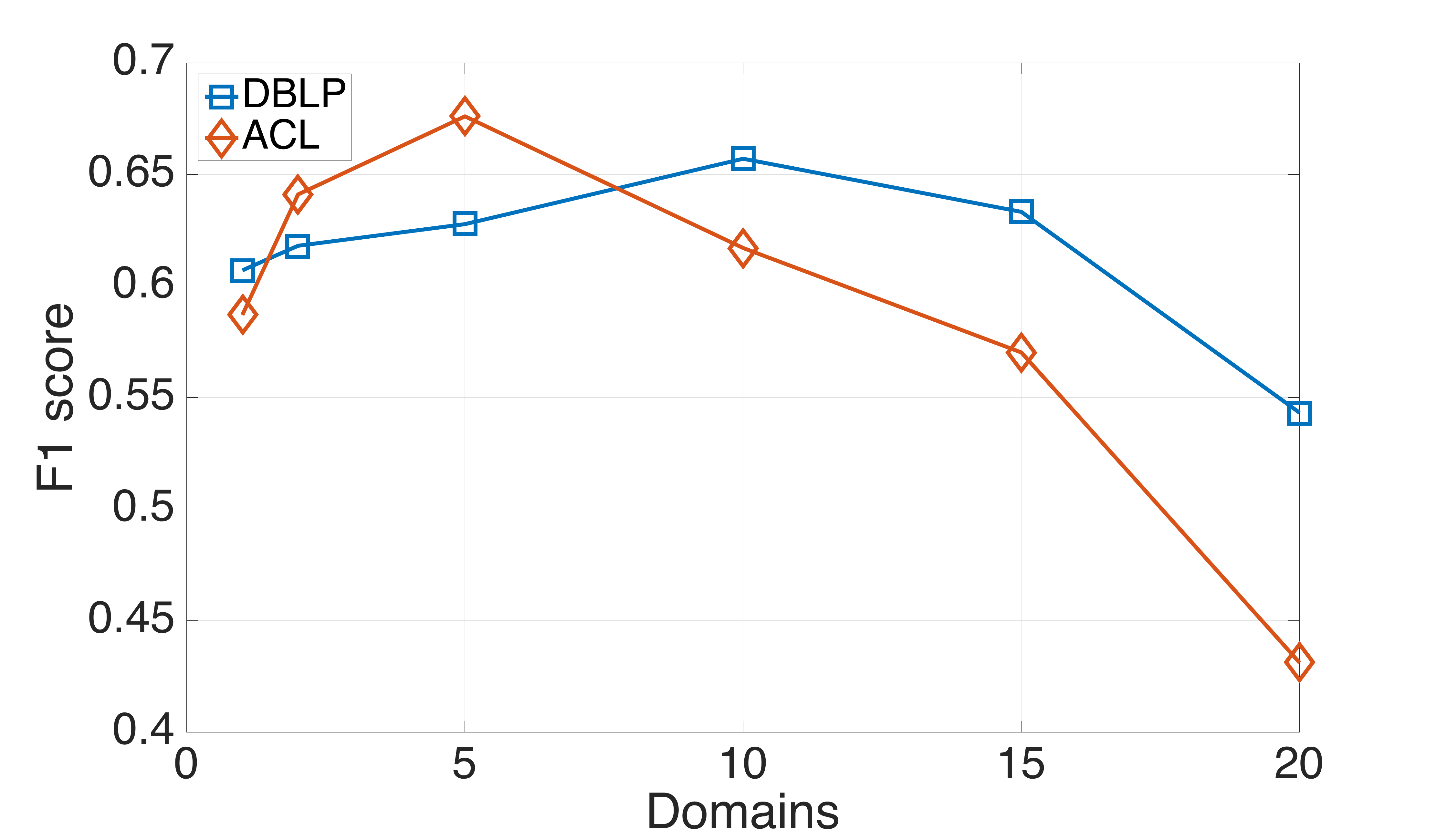}
        \label{performance_domains}}
\subfigure[b][]{
        \centering
        \includegraphics[width=0.32\linewidth]{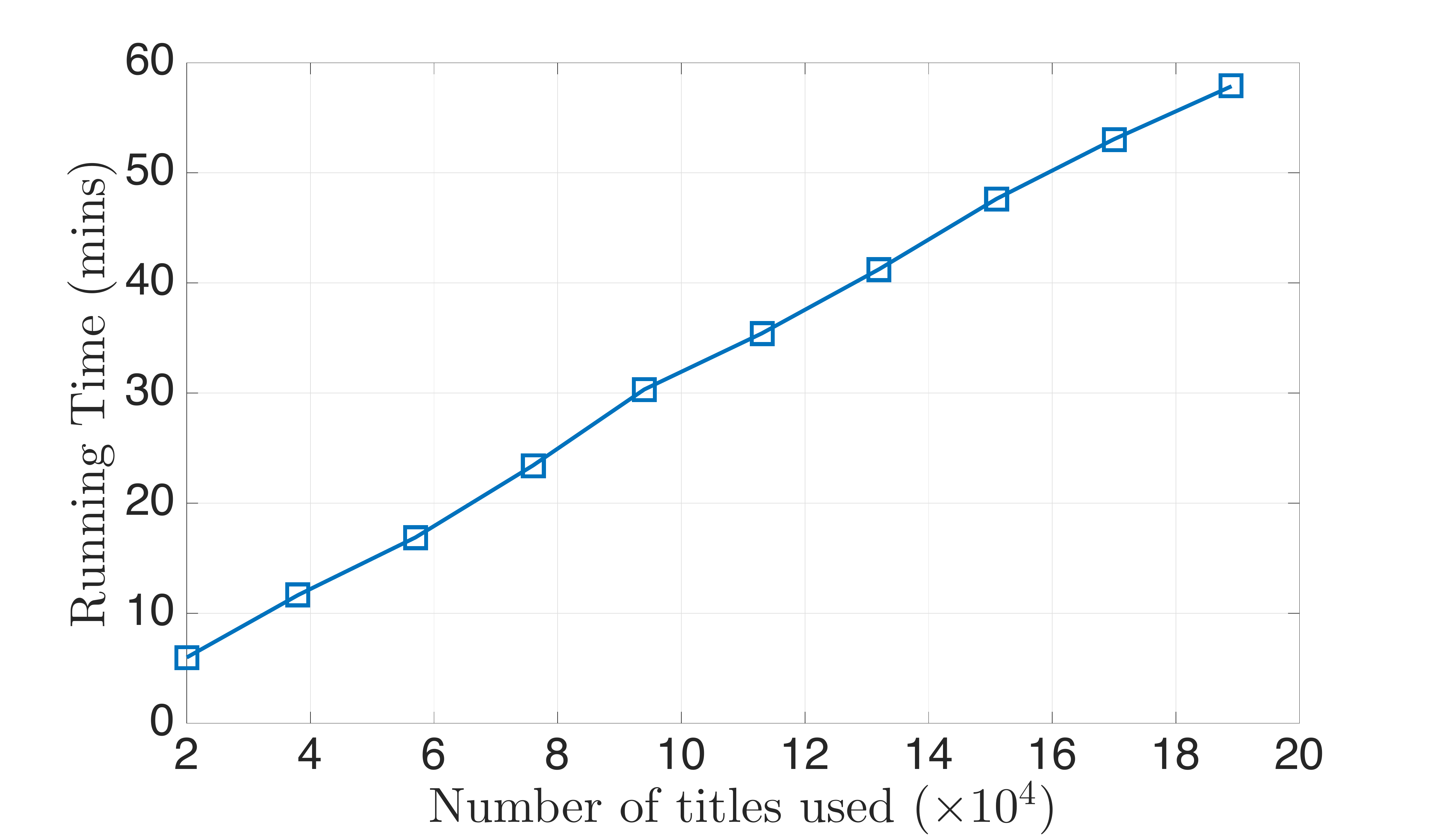}
        \label{runtime}}
      \vspace{-14.5pt}        
  \caption{ (a) Performance of \textit{DomainPhraseType} on varying the corpus size in DBLP dataset (b) Performance of \textit{DomainPhraseType} on varying the number of domains and (c) Runtime analysis for \textit{DomainPhraseType} on the 2 corpora}
\vspace{-2pt}
\end{figure*}

We find \textit{PhraseType + Adaptor} to outperform \textit{DomainPhraseType + Adaptor} by a small margin. \textit{PhraseType + Adaptor} is able to extract concepts of higher quality since it is learned on the entire corpus while \textit{DomainPhrase + Adaptor} performs concept extraction specific to each domain and could face sparsity in some domains, however this is offset by improved aspect typing by \textit{DomainPhraseType + Adaptor} in the identification of typed concept mentions. \vspace{5pt}\\\
\textbf{Identification of aspect-typed concept mentions:}
For aspect-typed concept mention identification, we first evaluate the performance of \textit{PhraseType + Adaptor} against the baselines on domain-specific subsets of DBLP (Table \ref{dblp_domains}). We then evaluate all techniques including \textit{DomainPhraseType + Adaptor/Adaptor:Mod} on the complete multi-domain ACL and DBLP datasets (Table \ref{dblp_aspect}). We find \textit{DomainPhraseType} based methods to outperform \textit{PhraseType} owing to improved aspect typing at the domain granularity.\vspace{5pt}\\\
\textbf{Effect of corpus size on performance:}
We vary the size of the DBLP dataset by randomly sampling a subset of the corpus in addition to the gold-standard annotated titles and measure the performance of different techniques (Fig \ref{performance_corpus}). We observe significant performance drop when the size of the corpus is reduced to $\leq$ 20\% of all titles, primarily due to reduced representation of sparse domains. Performance appears to be stable post 30\%.\vspace{5pt}\\
\textbf{Effect of number of domains:}
To observe the effect of number of domains on performance, we varied $|D|$ from 1 to 20 in the \textit{DomainPhraseType} model for the DBLP and ACL datasets as in Fig \ref{performance_domains}. Final results are reported based on the optimal number of domains, 10 for DBLP and 5 for ACL.\vspace{5pt}\\
\textbf{Runtime analysis:}
Our experiments were performed on an x64 machine with Intel(R) Xeon(R) CPU E5345 (2.33GHz) and 16 GB of memory. All models were implemented in C++. Our runtime was found to vary linearly with the corpus size(Fig \ref{runtime}). 
%
%

\begin{table}[t] \footnotesize
\vspace{10pt}
\begin{tabular}{@{}p{0.30\linewidth}p{0.65\linewidth}@{}}
\toprule
\textbf{Concept} & \textbf{Modifier} \\ 
\midrule
Approximation algorithm & Improved, Constant-Factor, Polynomial-Time, Stochastic, Distributed, Adaptive \\ 
Decision tree & Induction, Learning, Classifier, Algorithm, Cost-Sensitive, Pruning, Construction, Boosted\\ 
Wireless network & Multi-Hop, Heterogeneous, Ad-Hoc, Mobile, Multi-Channel, Large, Cooperative\\ 
Topic model & Probabilistic, Supervised, Latent, Approach, Hierarchical, LDA, Biterm, Statistical\\  
Neural network & Recurrent, Convolutional, Deep, Approach, Classifier, Architecture\\ 
Sentiment analysis & Aspect-Based, Cross-Lingual, Sentence-Level, In-Twitter, Unsupervised  \\
Image classification & Large-scale, Fine-grained, Hyperspectral, Multi-Label, Simultaneous, Supervised \\ \bottomrule
\end{tabular}
\caption{Modifiers for a few sample concepts}
\label{modifiers}
\vspace{-15pt}
\end{table}
\subsection{Case Studies}
\vspace{5pt}
\textbf{Top modifiers for sample concepts:}
We extract the modifiers obtained by \textit{DomainPhraseType +Adaptor} for a few sample concepts and depict the top modifiers (ranked by their popularity) in 
Table \ref{modifiers}. For a Technique concept such as \textit{Neural Network}, modifiers such as \textit{convolutional} and \textit{recurrent} represent multiple variations of the technique proposed in different scenarios. The modifiers extracted for a concept provide a holistic perspective of the different variations in which the particular concept has been observed in research literature.\vspace{5pt}\\
\textbf{Domains discovered in DBLP:}
In Table \ref{domains}, we provide a visualization of the domains  discovered by \textit{DomainPhraseType} in the DBLP dataset. Table \ref{domains} shows the 
the most probable venues ($\phi_v$) and a few popular concepts identified by \textit{DomainPhraseType + Adaptor} for the articles typed to each domain. An interesting observation is the ability of our framework to distinguish between fine-grained domains such as IR and NLP and identify the most relevant concepts for each domain accurately.

\begin{table*}[t] \small
\begin{tabular}{p{0.75in}ccccc}
\toprule
\textbf{Domain \#} & 1 & 2 & 3 & 4 & 5\\
\midrule
Top venues $\phi_{v}$ & SIGIR, CIKM, IJCAI & ICALP, FOCS, STOC	& OOPSLA, POPL, PLDI &	CVPR, ICPR, NIPS &	ACL, COLING, NAACL\\\hline
\multirow{3}{*}{Concepts} & web search	& complexity class &	flow analysis &	neural network &	machine translation \\ & knowledge base	& cellular automaton &	garbage collection &	face recognition	 & natural language \\ & search engine	& model checking	 & program analysis &	image segmentation &	dependency parsing\\ 
\midrule
\midrule
\textbf{Domain \#} & 6 & 7 & 8 & 9 & 10\\
\midrule
Top venues $\phi_{v}$ & ICDM, KDD, TKDE &	ICC, INFOCOM, LCN &	SIGMOD, ICDE, VLDB &	ISAAC, COCOON, FOCS&	WWW, ICIS, WSDM\\\hline
\multirow{3}{*}{Concepts} & feature selection&	sensor network&	database system& planar graph&	social network \\ & association rule &	cellular network	&data stream&  efficient algorithm&	information system \\ &time series &	resource allocation & query processing & spanning tree & semantic web\\ 
\bottomrule
\end{tabular}
\vspace{2pt}
\caption{Domains discovered by \textit{DomainPhraseType} in the DBLP dataset ($|D|$=10)}
\vspace{-16pt}
\label{domains}
\end{table*}
\section{Related Work}
The objective of our work is the automatic typing and extraction of concept mentions in short text such as paper titles, into aspects such as Technique and Application. Unlike typed entities in a traditional Named Entity Recognition(NER) setting such as people, organizations, places etc., academic concepts are not notable entity names that can be referenced from a knowledgebase or external resource. They exhibit variability in surface form and usage and evolve over time. Indicative features such as trigger words (Mr., Mrs. etc), grammar properties and predefined patterns are inconsistent or absent in most academic titles. Furthermore, NER techniques rely on rich contextual information and semantic structures of text \cite{nlp_entreg,wise}. Paper titles, on the other hand, are structured to be succinct, and lack context words. 

The problem of semantic class induction \cite{sci1, sci2} is related to typing of concept mentions since aspects are analogous to semantic classes. \cite{gn} studies the extraction of generalized names in the medical domain through a bootstrapping approach, however academic concepts are more ambiguous and hence harder to type. Many of them correspond to both Technique and Application aspects in different mentions, and hence must be typed in an instance specific manner rather than globally. To the best of our knowledge there has been very limited work in extraction of typed concept mentions from scientific titles or abstracts.

Phrase mining techniques such as \cite{topmine} and \cite{segphrase} study the extraction of significant phrases from large corpora, however they do not factor aspects or typing of phrases in the extraction process. We briefly summarize past approaches for academic concept extraction from the abstracts of articles. We also survey techniques that extract concept mentions within the full text of the article, which is not our primary focus. 

Concept typing has been studied in earlier work in the weakly supervised setting where Bootstrapping algorithms \cite{cikm13,ijcnlp} are applied to the abstracts of scientific articles, assuming the presence of a seed list of high-quality concept mention instances for each aspect of interest. \cite{ijcnlp} uses dependency parses of sentences to extract candidate mentions and applies a bootstrapping algorithm to extract three types of aspects -  focus, technique, and application domain. \cite{cikm13} uses noun-phrase chunking to extract concept mentions and local textual features to annotate concept mentions iteratively. Our experiments indicate that their performance is dependent on seeding domain-specific concepts. Furthermore, noun-phrase chunkers are dependent on annotated academic corpora for training. \cite{facetgist} extracts faceted concept mentions in the article text by exploiting several sources of information including the structure of the paper, sectional information, citation data and other textual features. However it is hard to quantify the importance of each extracted facet or entity mention to the overall contribution or purpose of the scientific article.

Topic models have also been recently used to study the popularity of research communities and evolution of topics over time in scientific literature \cite{evonetclus}. However, topic models that rely on statistical distributions over unigrams \cite{satm, dmm, cit1} do not produce sufficiently tight concept clusters in academic text. Citation based methods have also been used to analyze research trends \cite{citation}, however their key focus is understanding specific citations rather than extracting the associated concepts. Attribute mining \cite{am} combines entities and aspects (attributes) based on an underlying aspect hierarchy. Our work however identifies aspect-specific concept mentions at an instance level. \cite{sigir16} proposes an unsupervised approach based on pitman-yor grammars \cite{adaptor} to extract brand and product entities from shopping queries. However, brand and product roles are not interchangeable (a brand can never be a product) unlike academic concepts. Furthermore, most shopping queries are structured to place brands before product. Paper titles however are not uniformly ordered and thus need to be normalized by aspect typing their constituent phrases prior to concept extraction.
\vspace{-2pt}
\section{Conclusion}
In this paper, we address the problem of concept extraction and categorization in scientific literature. We propose an unsupervised, domain-independent two-step algorithm to type and extract key concept mentions into aspects of interest. PhraseType and DomainPhraseType leverage textual features and relation phrases to type phrases. This enables us to extract aspect and domain specfic concepts in a data-driven manner with adaptor grammars. While our focus here has been to apply our algorithm on scientific titles to discover technique and application aspects, there is potential to apply a similar two-step process in other domains such as medical text to discover aspects such as drugs, diseases, and symptoms. It is also possible to extend the models to sentences in full text documents while exploiting grammatic and syntactic structures. Our broader goal is to eliminate the need for human effort and supervision in domain-specific tasks such as ours.
\vspace{-3pt}
\section{Acknowledgments}
Research was sponsored in part by the U.S. Army Research Lab. under Cooperative Agreement No. W911NF-09-2-0053 (NSCTA), National Science Foundation IIS 16-18481 and NSF IIS 17-04532, and grant 1U54GM114838 awarded by NIGMS through funds provided by the trans-NIH Big Data to Knowledge (BD2K) initiative.
\vspace{-3pt}
\bibliographystyle{ACM-Reference-Format}
\bibliography{cikm_fp1860} 

\end{document}